\documentclass{article}
\usepackage{arxiv}
\usepackage{times}
\usepackage{array}
\usepackage{varwidth}
\usepackage{latexsym}
\usepackage{url}
\usepackage{lscape}
\usepackage{color,soul}
\usepackage{graphicx}
\usepackage{caption}
\usepackage{subcaption}
\usepackage[hidelinks]{hyperref}
\usepackage{longtable}
\usepackage{booktabs}
\usepackage{multirow}
\usepackage{amsmath}
\usepackage{authblk}
\usepackage[titletoc,title]{appendix}
\usepackage{lscape}
\usepackage{enumitem}
\usepackage[framed,numbered]{matlab-prettifier}

\usepackage[sorting=none]{biblatex} 
\usepackage{array}
\newcolumntype{L}{>{\centering\arraybackslash}m{2cm}}
\newcolumntype{W}{>{\centering\arraybackslash}m{2.5cm}}

\usepackage{floatpag}

\makeatletter
\def\namedlabel#1#2{\begingroup
    #2%
    \def\@currentlabel{#2}%
    \phantomsection\label{#1}\endgroup
}
\makeatother

\usepackage{microtype}
\usepackage{rotating}

\title{A systematic review of biologically-informed deep learning models for cancer: fundamental trends for encoding and interpreting oncology data}

\author[ ,1,2]{Magdalena Wysocka\thanks{Corresponding author: magdalena.wysocka@manchester.ac.uk}}
\author[ ,1,2]{Oskar Wysocki\thanks{The first two authors contributed equally.}}
\author[3]{Marie Zufferey}
\author[1]{D\'{o}nal Landers}
\author[1,2,3]{Andr\'{e} Freitas}

\affil[1]{digital Experimental Cancer Medicine Team, Cancer Biomarker Centre,\authorcr CRUK Manchester Institute, University of Manchester, Manchester, UK}
\affil[2]{Department of Computer Science, University of Manchester, Manchester, UK}
\affil[3]{Idiap Research Institute, Martigny, Switzerland}

\begin{document}

\date{}
\maketitle

\begin{abstract}

\textbf{Background}
There is an increasing interest in the use of Deep Learning (DL) based methods as a supporting analytical framework in oncology. However, most direct applications of DL will deliver models with limited transparency and explainability, which constrain their deployment in biomedical settings.

\textbf{Methods}
This systematic review discusses DL models used to support inference in cancer biology with a particular emphasis on multi-omics analysis. It focuses on how existing models address the need for better dialogue with prior knowledge, biological plausibility and interpretability, fundamental properties in the biomedical domain. For this, we retrieved and analyzed 42 studies focusing on emerging architectural and methodological advances, the encoding of biological domain knowledge and the integration of explainability methods. 

\textbf{Results}
We discuss the recent evolutionary arch of DL models in the direction of integrating prior biological relational and network knowledge to support better generalisation (e.g. pathways or Protein-Protein-Interaction networks) and interpretability. This represents a fundamental functional shift towards models which can integrate mechanistic and statistical inference aspects. We introduce a concept of \textit{bio-centric interpretability} and according to its taxonomy, we discuss representational methodologies for the integration of domain prior knowledge in such models. 

\textbf{Conclusions}
The paper provides a critical outlook into contemporary methods for explainability and interpretabiltiy used in DL for cancer. The analysis points in the direction of a convergence between encoding prior knowledge and improved interpretability. We introduce \textit{bio-centric interpretability} which is an important step towards formalisation of \textit{biological interpretability} of DL models and developing methods that are less problem- or application-specific. 

\end{abstract}

\section{Introduction}

There is an increasing interest in the use of Deep Learning (DL) based methods as a supporting analytical framework in oncology. Recent work have articulated the potential applied impact of DL-based methods in oncology including drug response prediction \cite{10.1093/bib/bbz171, Sharifi-Noghabi2019} or combination \cite{10.1214/20-STS783}, cancer diagnosis or prognosis \cite{kumarSystematicReviewArtificial2021, tranDeepLearningCancer2021, tufailDeepLearningCancer2021, jiao2020, hassanzadehIntegratedDeepNetwork2021, Kipkogei2021.10.11.21264761} and the overall impact of this emerging analytical substrate to deliver the vision of precision and personalised medicine \cite{bhinderArtificialIntelligenceCancer2021, tranDeepLearningCancer2021}. Despite not being mainstream methods at this point, these architectures point in the direction of addressing existing paradigmatic analytical gaps currently faced by more traditional inference frameworks, including the tension between small study cohorts and increasingly available complex set of features per patient ($p >> n$). 

However, most direct applications of DL will deliver models with limited transparency and explainability, which constrain their deployment in biomedical settings. In this systematic analysis we tackle an aspect commonly acknowledged but left almost untouched, namely: how authors understand and use the definition of biological interpretability, and how it dialogues to the growing spectrum of biologically-informed models, which integrate prior biological knowledge within existing DL frameworks. 
This paper provides a systematic review focused on omics-based DL models used in cancer biology \textit{highlighting the dialogue and convergence between biologically-informed models, explainable AI (XAI) and biological interpretability}. We perform a systematic review, identifying the motifs within emerging architectures:  the domain knowledge which is integrated in the design of the models, data representation aspects and emerging architectures, ranging from biological networks and graphs to embedding models. We introduce the concept of {\textit{bio-centric} interpretability} in DL models, which augments the contemporary Explainable (XAI) taxonomies and emerges as a fundamental property and desideratum of biologically-informed DL.

Addressing the above mentioned gaps, we defined the following research questions:

\begin{enumerate}
    \item What are the perspectives of interpretability accross different DL-based frameworks within the cancer research domain?
    
    \item What are the methods that deliver biological interpretability?
    
    \item What are the desirable approaches to integration of domain knowledge in the models' architecture?
    
    \item What are the emerging representation paradigms within these models?

\end{enumerate}

\begin{figure}[ht!]
\centering
\includegraphics[width= .6\textwidth]{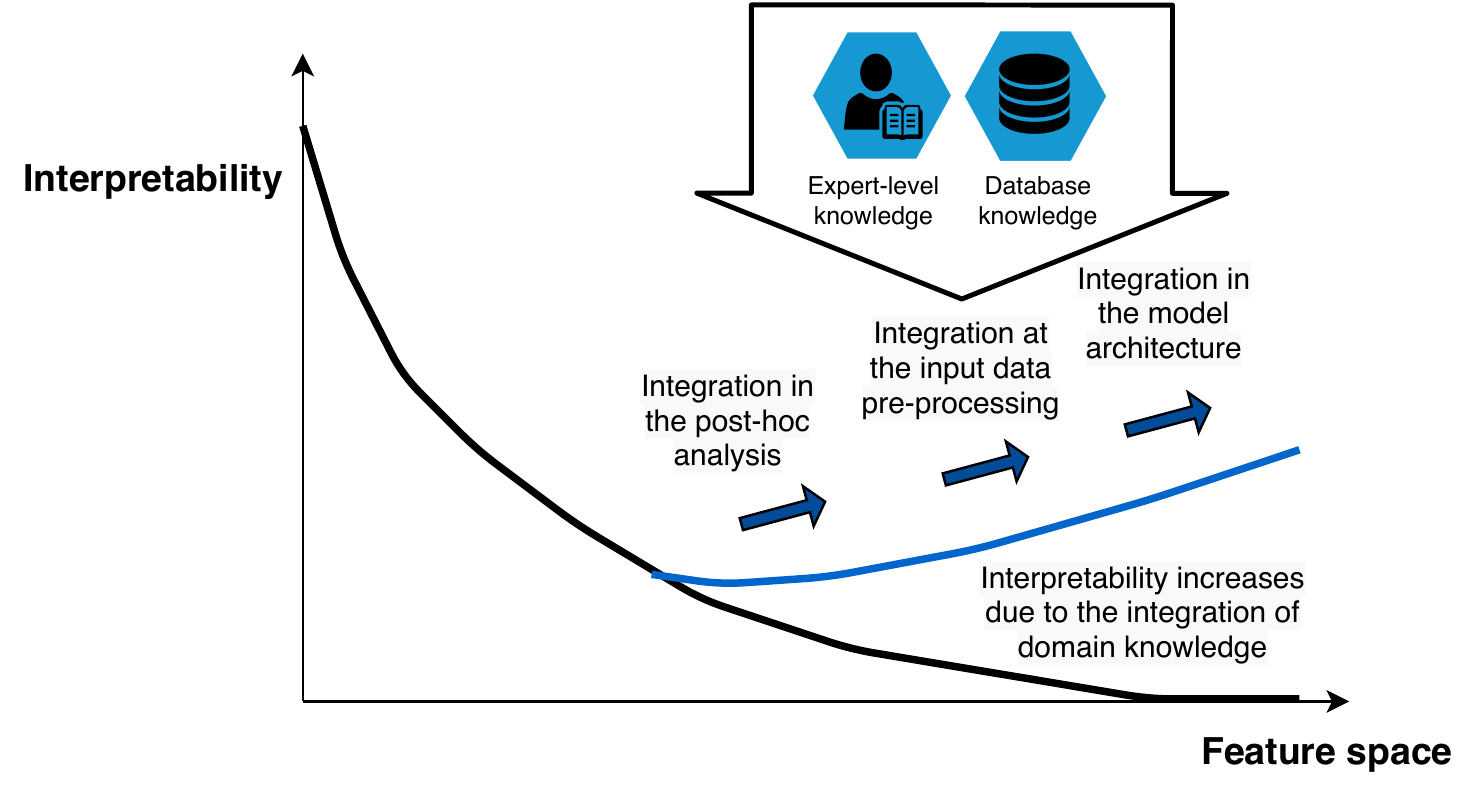}
\caption{Current mechanisms to restore or increase biological interpretability
There is a negative correlation between the interpretability of the DL model and the size of the feature space. However, the integration of domain knowledge can systematically address this dimensionality issue. Domain knowledge can be distinguished between expert-level knowledge and knowledge derived from databases. The motifs for integrating domain knowledge within emerging architectures were identified: the domain knowledge which a) is integrated in the design of the models, b) is integrated in the input data pre-processing, c) is integrated in the post-hoc analysis process.
}
\label{fig:fig_1}
\end{figure}

In addition to recent surveys in Explainable AI (XAI) (i.a. \cite{MONTAVON20181, DBLP:journals/access/AdadiB18, 10.1145/3236009, Marcinkevics2020, 10.3389/fdata.2021.688969, samekExplainingDeepNeural2021, DBLP:journals/corr/abs-2010-00389}), XAI in the field of genomics \cite{10.1093/bib/bbaa177, watson2021interpretable, wysocki2022transformers} and medicine \cite{holzinger2017need, doi:10.1098/rsif.2017.0387, stiglicInterpretabilityMachineLearning2020, Tjoa2021ASO, yangDysregulationBrainChoroid2021, coronetuserstudywysocki}, we highlight a much more specific subfield, aiming to link explainability and biological interpretability. The systematic review is restricted to the context of multi-omics based DL in cancer biology, excluding papers from the computer-vision subarea. We focus on the dialogue between post-hoc explainability (regarding its internal mechanisms and the interpretation of the model's output) and the encoding of prior biomedical knowledge, thus discussing the contribution of AI for supporting the understanding of oncogenic processes, in particular, the methods for integrating existing domain knowledge (DK) into DL models. 
We highlight the dialogue between explicit and latent representations.
A diagrammatic outline of the discussion is depicted in Fig. \ref{fig:fig_1}.

\section{Results}

The electronic bibliographic databases (PubMed and Web of Science) search identified 661 records, which were reduced to 591 after removing duplicates. The 591 records were screened on the basis of prespecified inclusion criteria resulting in 176 records. All these potentially relevant articles were read in full text. The reasons for the exclusion of the papers were as follows: papers provided methods that are not directly linked to cancer and functional analysis/insights on biological processes; papers provided models based on DL and ML using clinical/laboratory data alone; based on microarray data or developed a sequence-based algorithmic framework. A list of eligible studies was created and resulted in 42 studies\footnote{Each of the selected studies is from the existing state of the art and not performed by any of the authors.} (Table 1).

\fontsize{6}{11}\selectfont{
{\setlength{\tabcolsep}{1.4pt}
\begin{longtable}[c]{@{}ccLLLLLLL@{}}
\caption{Strategies of domain knowledge integration and explainability methods}
\label{tab:summary_table}\\
\toprule
Reference &
  Year &
  Type of omic input data &
  Model's task &
  Domain knowledge &
  Domain knowledge database &
  DK integration (PRE - input data preprocessing, ARCH - model's architecture, POSTHOC) &
  DL model type / DL architecture &
  Explainability method used \\* \midrule
\endfirsthead
\multicolumn{9}{c}%
{{\bfseries Table \thetable\ continued from previous page}} \\
\toprule
Reference &
  Year &
  Type of omic data &
  Model's output &
  Prior knowledge &
  Prior knowledge database &
  DK integration (PRE - input data preprocessing, ARCH - model's architecture, POSTHOC) &
  DL model type / DL architecture &
  Interpretability method used \\* \midrule
\endhead
\bottomrule
\endfoot
\endlastfoot
\cite{chingCoxnnetArtificialNeural2018} &
  2018 &
  single omic: T &
  survival prediction &
  pathways &
  KEGG &
  POSTHOC &
  DNN &
  partial derivatives \\
\cite{haoPASNetPathwayassociatedSparse2018} &
  2019 &
  single-omic: T &
  survival classification &
  pathways &
  MSigDB (Reactome) &
  ARCH &
  sparse DNN &
  weights and nodes activation \\
\cite{kinalisDeconvolutionAutoencodersLearn2019} &
  2019 &
  sc single-omic: T &
  biological functions determination &
  pathways &
  DMAP &
  POSTHOC &
  AE &
  saliency maps \\
\cite{lemsaraPathMEPathwayBased2020} &
  2020 &
  multi-omics: T, E, G &
  patient stratification &
  pathways &
  NCI: PID &
  PRE &
  sparse DAE &
  SHAP \\
\cite{guoDeepLearningbasedOvarian2020} &
  2020 &
  multi-omics: T, G &
  patient stratification &
  pathways &
  KEGG &
  POSTHOC &
  DAE &
  latent space exploration \\
\cite{elmarakebyBiologicallyInformedDeep2021} &
  2021 &
  multi-omics: T, G &
  predict disease state &
  pathways &
  Reactome &
  ARCH &
  sparse DNN &
  DeepLIFT \\
\cite{fengInvestigatingRelevanceMajor2021} &
  2021 &
  multi-omics: T, G + clinical data &
  survival prediction &
  pathways &
  KEGG &
  ARCH &
  sparse DNN + CNN &
  smoothgrad \\
\cite{Zhang2021} &
  2021 &
  multi-omics: T, G &
  drug response prediction &
  pathways &
  KEGG &
  ARCH &
  sparse DNN &
  smoothgrad \\
\cite{ohPathCNNInterpretableConvolutional2021} &
  2021 &
  multi-omics: T, E, G + clinical data &
  survival classification &
  pathways &
  KEGG &
  PRE &
  CNN &
  Grad-CAM \\
\cite{zhaoDeepOmixScalableInterpretable2021} &
  2021 &
  multi-omics: T, E, G &
  survival prediction &
  pathways (to define functional modules) &
  MSigDB (KEGG and Reactome) &
  ARCH &
  sparse DNN &
  functional module activity \\
\cite{fortelnyKnowledgeprimedNeuralNetworks2020} &
  2020 &
  sc single-omic: T &
  molecular interpretation of single-cell RNA-seq data &
  pathways and gene regulatory networks &
  SIGNOR, Harmonizome, TTRUST &
  ARCH &
  sparse DNN &
  in silico perturbation \\
\cite{withnellXOmiVAEInterpretableDeep2021} &
  2021 &
  single-omic: T &
  cancer type classification &
  pathways, Gene Ontology &
  Reactome, GO &
  POSTHOC &
  VAE &
  DeepSHAP \\
\cite{dengPathwayGuidedDeepNeural2020} &
  2020 &
  single-omic: T + drug targets + drug sensitivity &
  drug response prediction &
  pathways; drug sensitivity; drug-protein interactions &
  KEGG, GDSC, STITCH &
  ARCH &
  sparse DNN &
  pathway node activity \\
\cite{seningeVEGAInterpretableGenerative2021} &
  2021 &
  sc single-omic: T &
  inferring gene module activities &
  pathways; gene regulatory networks; cell type identities (to define gene modules) &
  MSigDB (mainly Reactome and Hallmark) &
  ARCH &
  sparse VAE &
  latent space exploration \\
\cite{leeCancerSubtypeClassification2020} &
  2020 &
  single-omic: T &
  cancer type classification &
  pathways; PPI network &
  KEGG, BioGRID &
  PRE+ARCH +POSTHOC &
  GCN-MLP &
  attention mechanisms \\
\cite{chuangConvolutionalNeuralNetwork2021} &
  2021 &
  single-omic: T &
  cancer type classification &
  PPI network &
  BioGRID, DIP, InAct, MINT, MIPs &
  PRE &
  CNN &
  NA \\
\cite{cheredaExplainingDecisionsGraph2021} &
  2021 &
  single-omic: T &
  metastatic or non-metastatic classification &
  PPI network &
  HPRD &
  PRE &
  GCN &
  GLRP \\
\cite{liuNetworkbasedDeepLearning2021} &
  2021 &
  single-omic: G + clinical data &
  patient stratification &
  PPI network &
  \textgreater 10 PPI databases &
  PRE &
  struc2vec &
  latent space exploration \\
\cite{althubaitiDeepMOCCAPancancerPrognostic2021} &
  2021 &
  multi-omics: T, E, G + clinical data &
  survival prediction &
  PPI network; driver genes &
  STRING, COSMIC &
  PRE &
  GCN &
  attention mechanisms; latent space exploration \\
\cite{ramirezClassificationCancerTypes2020} &
  2021 &
  single-omic: T &
  cancer type classification &
  PPI network; other (co-expression graph) &
  STRING &
  PRE &
  GCN &
  in silico perturbation \\
\cite{Schulte-Sasse2019} &
  2021 &
  multi-omics: T, E, G &
  cancer gene prediction &
  PPI network; other (Gene-Gene interaction network, network of cancer genes NCG) &
  CPDB; collecion of labels from NCG, COSMIC, OMIM, KEGG &
  PRE+POSTHOC &
  GCN &
  LRP \\
\cite{liuTranSynergyMechanismdrivenInterpretable2021} &
  2021 &
  single-omic: T &
  drug response prediction &
  drug-target interaction, PPI network &
  DrugBank, ChEMBL, STRING &
  PRE &
  transformer + GCN &
  Shapley Additive Gene Set Enrichment Analysis (SAGSEA) \\
\cite{chiuPredictingCharacterizingCancer2021} &
  2021 &
  multi-omics: T, E, G &
  gene dependency / cancer dependencies &
  fingerprint vector &
  MSigDB &
  PRE &
  AE+DNN &
  in silico perturbation \\
\cite{Ma2018} &
  2019 &
  single-omic: T &
  patient stratification &
  Gene Ontology &
  GO &
  ARCH &
  factor GNN &
  attention mechanisms; RLIPP \\
\cite{kuenziPredictingDrugResponse2020} &
  2020 &
  single-omic: T + chemical structure network &
  drug response prediction &
  Gene Ontology &
  GO &
  ARCH &
  DNN + sparse DNN &
  RLIPP \\
\cite{huangSALMONSurvivalAnalysis2019} &
  2019 &
  multi-omics: T+T + clinical data (not mandatory) &
  survival prediction &
  other (co-expression modules from lmQCM algorithm) &
  NA &
  PRE +POSTHOC &
  DNN &
  in silico perturbation \\
\cite{xingMultiLevelAttentionGraph2020} &
  2020 &
  single-omic: T &
  survival outcome, histological grading &
  other (format the omic data into WGCNA graphs) &
  NA &
  PRE &
  GNN &
  full-gradient graph saliency (FGS) \\
\cite{caoMultiomicsIntegrationRegulatory2021} &
  2021 &
  sc multi-omics: T, E, G &
  regulatory process &
  other (knowledge-based graph “guidance graph”) &
  NA &
  PRE &
  VAE &
  latent space exploration \\
\cite{gao2020} &
  2020 &
  multi-omics: T, G + clinical data &
  survival classification &
  other (learnt bipartite graphs) &
  NA &
  PRE &
  GNN &
  NA \\
\cite{shuModelingGeneRegulatory2021} &
  2021 &
  sc single-omic: T &
  single-cell computational tasks &
  other (learnt gene regulatory network) &
  NA &
  ARCH &
  VAE &
  latent space exploration \\
\cite{wangMOGONETIntegratesMultiomics2021} &
  2021 &
  multi-omics: T, E &
  cancer type classification &
  other (sample similarity networks) &
  NA &
  PRE &
  GCN + VCDN &
  in silico perturbation \\
\cite{Way174474} &
  2018 &
  single-omic: T &
  cancer type classification &
  NA &
  NA &
  NA &
  VAE &
  latent space exploration \\
\cite{Titus433763} &
  2018 &
  single-omic: E &
  cancer type classification &
  NA &
  NA &
  NA &
  VAE &
  latent space exploration \\
\cite{Sharifi-Noghabi2019} &
  2019 &
  multi-omics: T, G &
  drug response prediction &
  NA &
  NA &
  NA &
  DNN &
  NA \\
\cite{simidjievskiVariationalAutoencodersCancer2019} &
  2019 &
  multi-omics: T, G &
  patient stratification &
  NA &
  NA &
  NA &
  VAE &
  latent space exploration \\
\cite{10.1093/bioinformatics/btz158} &
  2019 &
  single-omic: T &
  drug response prediction &
  NA &
  NA &
  NA &
  VAE &
  latent space exploration \\
\cite{wangExtractingBiologicallyLatent2019} &
  2019 &
  single-omic: E &
  cancer type classification &
  NA &
  NA &
  NA &
  VAE &
  latent space exploration \\
\cite{palazzoPancancerSomaticMutation2019} &
  2019 &
  single-omic: G &
  cancer type classification &
  NA &
  NA &
  NA &
  AE &
  latent space exploration \\
\cite{jiao2020} &
  2020 &
  single-omic: G &
  cancer type classification &
  NA &
  NA &
  NA &
  DNN &
  NA \\
\cite{hiraIntegratedMultiomicsAnalysis2021} &
  2021 &
  multi-omics: T, E, G &
  survival classification &
  NA &
  NA &
  NA &
  VAE &
  latent space exploration \\
\cite{hassanzadehIntegratedDeepNetwork2021} &
  2021 &
  multi-omics: T, E for KIRC or multi-omics: T+T for HNSC &
  survival prediction &
  NA &
  NA &
  NA &
  DBN &
  NA \\
\cite{Kipkogei2021.10.11.21264761} &
  2021 &
  multi-omics &
  survival classification &
  NA &
  NA &
  NA &
  transformer &
  attention weights \\* \bottomrule
\end{longtable}
Legends: Types of omic data: G - genomics, P - proteomics, T - transcriptomics, E - epigenomics; GO - Gene Ontology; PPI - protein-protein interaction; WGCNA - Weighted Correlation Network Analysis; Deep Learning (DL) architecture: AE - autoencoder, ANN - Artificial Neural Networks, CNN - Convolutional Neural Network, DAE - Denoising Autoencoder, DBN - Deep Belief Network, DNN - Deep Neural Network, GCNN - graph convolutional neural network, GCNN-MLP - GCNN multilayer perceptron, MMD-VAE - Maximum Mean Discrepancy Variational Autoencoder, VAE - Variational Autoencoder, VCDN - View Correlation Discovery Network; Interpretability method: LRP - layer-wise relevance propagation; Interpretability group: II - intrinsically interpretable (Simulatability, Decompasability, Algorithmic transparency), PH - post-hoc (textual explanation, visualisation, local explanation, explanation by example); Interpretability group: PROC - Processing; REPR - Representation; CREATE - Explanation producing; NA - not applicable\\
}
}
\normalsize

\subsection{Emerging methodological paradigm: bio-centric model interpretability}

Explainability and interpretability are considered as key desiderata of the machine learning (ML) models (e.g. \cite{Marcinkevics2020}). They are thought to prevent the risks of misuse of machine learning models embedded in healthcare applications. Model transparency and explainability are required to deploy AI-derived biomarkers in clinical settings. In addition, the transparency of interpretable methods can minimise the risks in AI-based decision-making in healthcare applications. It is by definition impossible to appeal to decisions resulting from a DL model that are not presented in an understandable manner and cannot be explained in biomedical terms and grounded in current biomedical reasoning. In biomedicine, the predictions and metrics calculated from these predictions alone are insufficient to characterise the model.

The existence of multiple types and definitions of models' interpretability makes it difficult to formulate a precise definition of \textit{biological interpretability} in a cancer biology setting. When is it valid to say that the ML model used in cancer biology is interpretable? The lack of a formal definition needs to be addressed and points in the direction of an unmet research gap. Benk and Ferrario \cite{benkExplainingInterpretableMachine2020} introduced three different dimensions of the need for interpretation: epistemic, pragmatic, ethical. In biology, the impact of these models from a scientific epistemology setting needs to be considered as, at their limit, emerging AI methods bring the promise of integrating heterogeneous evidence and mechanistic and statistical inference paradigms. These methods can ultimately impact fundamental notions of what constitutes a valid scientific argument, bringing alternative perspectives to the notion of statistical significance.

Despite high demand, interpretability remains one of the biggest challenges for bringing these models into a real-world setting. In the AI and ML fields, there is a well-known trade-off between how well the model performs and how well people are able to interpret it \cite{liptonMythosModelInterpretability2017, benkExplainingInterpretableMachine2020, doshivelez2017rigorous}. Additionally, there is no consistent agreement on definitions of interpretability. One of its definitions directly refers to the components of interpretable models such as transparency (`how does the model work?') and post-hoc explanations (`what else can the model tell me?') \cite{liptonMythosModelInterpretability2017}. It identifies two main objects for interpretation: i) the internal mechanisms, i.e. how the models compute their outcomes, and ii) the outcomes generated by the model. Similarly, according to known taxonomic accounts \cite{stiglicInterpretabilityMachineLearning2020}, interpretability can be: \textit{algorithmic-centric}, focusing on the inner-working of the model; or \textit{output-centric}, highlighting the model agnostic post-hoc analysis. 

In the context of DL, we replace algorithmic-centric with architecture-centric interpretability. We argue that more emphasis and inference is put on the structure of the model rather than the learning process of the DNN (via backpropagation algorithm).

In order to derive biological insights from the model, an interpretation of a biological expert is required regardless of architecture or outcome-centric approach. Both of them need to favor mapping the biological mechanisms to the models' components, aiming at delivering an interpretation for the intended end user (i.e. biologist, oncologist) which relies more on biological knowledge rather than on DL or mathematical knowledge. More specifically, a preferable format of model's transparency would be a biological mechanism integrated in the model's architecture (e.g. gene activation pathway) or calculations mimicking biological processes (e.g. mimic typical molecular biology assays that study functional genomics), in complement to state-of-the-art explanation methods borrowed from other fields. Some formats have already been successfully applied to transcriptomic data, such as the integration of DK of gene modules, or the integration of hierarchical information about molecular subsystems involved in cellular processes. Such models provide informative biological interpretation of the predictions by studying the activation of the various subsystems embedded in the model architecture and, moreover, they can make it possible to infer on the activity of latent factors as a priori characterized gene modules. The interpretation of the biological expert allows for evaluation of biological plausibility and satisfiability of biological constrains.

Hence, in this paper we revisit the notion of interpretability to ground it in a biomedical context, introducing the concept \textbf{\textit{bio-centric interpretability}}. It encompasses three key aspects which lead to biological understanding of the investigated problem and new insights:
\begin{itemize}
    \item architecture-centric interpretability
    \item output-centric interpretability
    \item post-hoc evaluation of biological plausibility
\end{itemize}

We argue, that evaluation of the DL model regarding bio-centric interpretability requires an analysis of all these aspects at once. 
These three aspects are evaluated via the analysis of the four bio-centric interpretability components: 
\begin{itemize}
    \item the integration of different data modalities
    \item the schema level representation of the model
    \item the integration of domain knowledge 
    \item post-hoc explainability methods
\end{itemize}

\subsubsection{The integration of different data modalities.}
Cancer is a complex and multi-faceted disease with a landscape of features that can separately or together influence treatment responses and patient prognosis.
Important biological relations can be expressed in more than one data modality, e.g. potential cancer driver genes can be represented through integration of copy number, DNA methylation and gene expression data. 
Therefore, combining different data modalities in the DL model, including different types of omics data is imminent as the field evolves and inherent if biological processes are modelled.
Only provided that the biologically-informed model can reveal both established and novel molecularly altered candidates which can be implicated in predicting advanced disease.

\subsubsection{Schema level representation of the model.} 
Understanding the data flow in the model is crucial for the post-hoc interpretation by an expert user. Obviously, this is affected by how the data is represented in subsequent components of the model. Usually, collected multi-omics data is stored in tables (matrices). However, over a series of computations steps, the representation can change into graphs, networks, eigenvalues, eigenvectors, among others. Each representation has its own specific properties and is processed by specific architectural elements in the model, e.g. Graph Neural Networks and Graph Convolution Networks (for graphs). Thus, in the context of bio-centric interpretability, it is crucial to understand these representations, how they transform and how to communicate such transformation during the post-hoc inference. The underlying dialogue between the input data model and the architectural structure of the model requires a schema level representation, which then allows for domain expert interpretation and inference. 

\subsubsection{The integration of domain knowledge.} 
A key aspect, which significantly impacts all three components is the domain knowledge integration \textit{into} the model. A biologically-informed DL model can and should make use of databases that contain an abundance of known biological relations. Later in the paper, we compare and contrast emerging approaches of DK integration and its close dialogue with DL archtectures, indicating which model has the highest potential in improving bio-centric interpretability. 

\subsubsection{Post-hoc explainability methods.} 
The inherent property of DL model is its ability to derive latent features reflected in a large space of weighted connections between neurons. Even provided that the model's architecture resembles bio relations, post-hoc explainability methods must be applied to allow for tracking back the information flow, highlighting the importance (and unimportance) of model's components. More specifically, when investigating an individual output it is necessary to define key neurons, connections or layers that most impact the prediction, as well as those that do not. Four areas of the bio-centric interpretability.
 
\subsection{Encoding domain priors: Improving bio-centric interpretability and integrating relational knowledge}

In cancer, AI / ML is emerging as a methodological enabler to transform omics data into biomarker panels that can diagnose, predict or report on the effectiveness of interventions in the disease. More recently, some of these methods have concentrated on the integration of symbolic-level, explicit domain knowledge into the models. Domain knowledge can be understood as the information so far accumulated in a given field (here: pathways, PPI networks, Gene Ontology), usually expressed as known relational knowledge. In many cases, this knowledge is available in well-known curated databases and expressed in canonical data models that can be integrated in a computational pipeline. The taxonomy for explicit knowledge integration with the informed ML framework proposed by von Rueden et al. \cite{vonruedenInformedMachineLearning2021} includes: i) source of knowledge; ii) representation of knowledge; iii) and integration of knowledge in the ML pipeline. Each dimension contains a set of elements showing different approaches that can be observed in previous literature. Knowledge sources can be classified according to the degree of formality. They range from the rigorously expressed scientific knowledge (derived from any scientific discipline) to an expert-derived statement (mapping for example their clinical experience). More or less formalised, more general scientific knowledge (aka. world knowledge) situated at a basic expertise level within that domain (e.g. that the body is composed of cells; that there is DNA inside cell nucleus; that cancer is a disease of the genome, etc.); we found the general scientific knowledge not relevant in the context of this work. 

Domain knowledge can be integrated into the model to improve its consistency, reliability and biological plausibility as well as for supporting better generalisation. As proposed by von Rueden et al. \cite{vonruedenInformedMachineLearning2021} this can be done in a variety of ways, such as incorporating DK into basic training data (e.g. pre-processing), hypothesis set (e.g. sparse connections between neurons), established relational data, learning algorithm (e.g. cost function), and final hypothesis (e.g. model's architecture). On the other hand, DK is needed in order to extract the scientific outcome from the model or from individual elements of the model, and/or to explain such outcome. For example, based on DK, the contributions of specific model components can be better localised and investigated. 

In addition, DK can be used in a post-hoc setting, where the scientific credibility and consistency of the results are cross-validated within existing knowledge. Results that do not match the existing knowledge can be rejected or flagged as incorrect or suspicious, so that the final result is consistent with prior knowledge. 

In this paper we define a taxonomy which is more specific for biologically-informed DL models (inspired by von Rueden \cite{vonruedenInformedMachineLearning2021}). We suggest three main categories of DK integration as:

\begin{itemize}
    \item input data pre-processing (PRE) - DK is used to enrich or augment the input data, which results in a change of data representation. Scaling or normalisation is excluded from this category.
    \item architecture definition (ARCH) - DK explicitly impacts the model architecture, such as connections between neurons and layers. 
    \item post-hoc comparison (POST-HOC) - DK is used to investigate and explain the outcome of the model. The DK is used to process the outcome and compare to current, known biological relations.
\end{itemize}

Multiple types of DK integration can be observed in a single model.

Of note, a pre-requisite of developing any DL model in cancer biology is to understand the target domain, needed at least to define the input and output, and to qualitatively or quantitatively evaluate this output. Despite acknowledging the expert knowledge of the authors of the models, we do not consider it as explicitly integrated domain knowledge. We consider the post-hoc DK integration when the output is compared with information derived from external knowledge, or the representation of the output is changed (e.g. a vector to a graph) by using DK, so the the biological plausibility can be validated.

\begin{figure}[ht!]
\centering
\includegraphics[width= .6\textwidth]{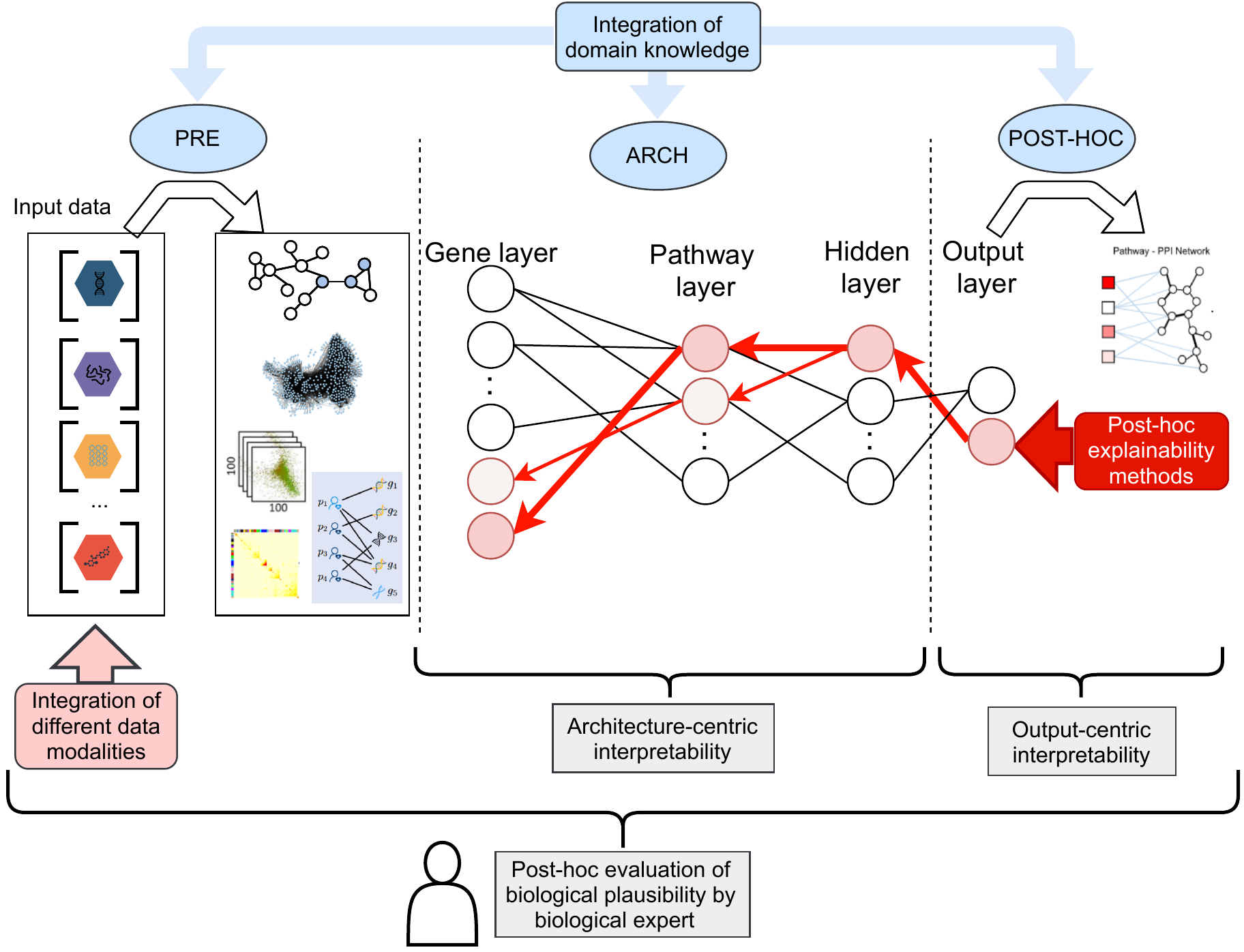}
\caption{Bio-centric interpretability scheme in the overview of a biologically-informed DL model. Grey boxes - three interpretability components
}
\label{fig:overview_biocentric_DL}
\end{figure}

An outline of the three categories of DK integration is shown in Fig. \ref{fig:overview_biocentric_DL}.
The results for selected papers according to proposed taxonomy are summarized in Table \ref{tab:summary_table}.

\subsection{Trends in DL models for cancer}

\begin{figure}
\centering
\includegraphics[width= 0.8\textwidth]{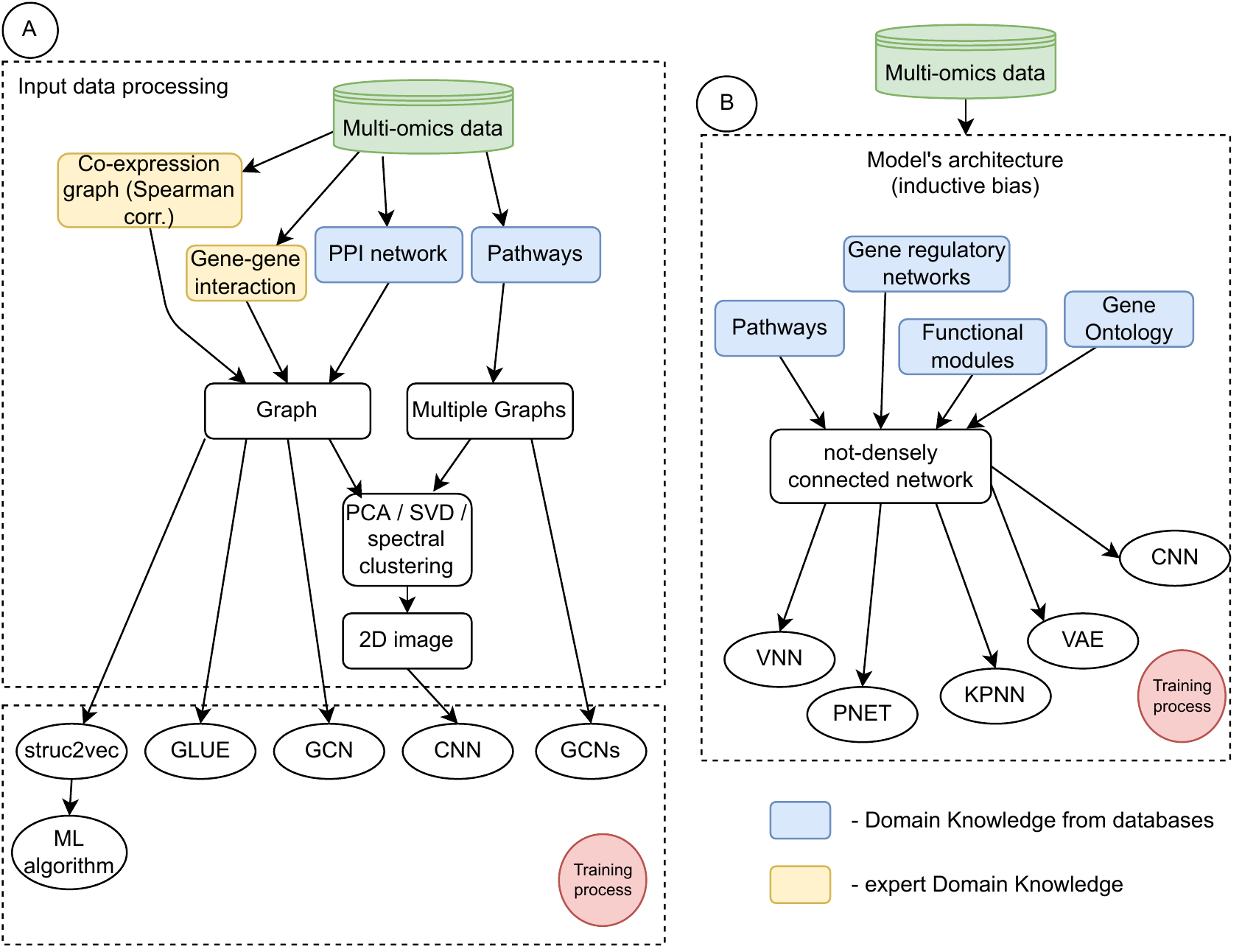}
\caption{Data representation paradigms and the impact of the integration of domain knowledge
Domain knowledge (DK) can be derived from a database (blue blocks) or expert DK (yellow blocks). DK can be used in pre-processing and data augmentation before the training process. DK from databases can be represented in two ways: \textbf{A} as a step in the pre-processing of input data, before the training process. This first paradigm has emerged for the representation of multi-omic data, which are transformed into graphs or a network and fed into GNN or GCN. This paradigm has been applied to DL models such as: struc2vec, GLUE, several GCN and CNN models; \textbf{B} as inductive bias when creating the neural network architecture, defining the connections between nodes in layers. In this case, DK impacts the training process as it affects the back-propagation. This paradigm has emerged mainly for the representation of multi-omic data, which are fed into sparsely connected Deep Neural Network, where connections are defined by biological relations. This paradigm has been applied to DL models such as: VNN, PNET, KPNN, VAE, CNN. 
}
\label{fig:DK_integration_diagram}
\end{figure}

The prominent explanation for the high heterogeneity observed in cancer may be the organisation of genes in various signalling/regulatory pathways and protein complexes. Cellular-level processes and responses are carried out by spatially and temporally organized sets of interacting entities such as proteins or RNA molecules. It is fundamental to understand how these interactions lead to biological processes. The conventional approach to studying biological processes is based on molecular interaction networks between individual biological molecules, represented as nodes with edges describing the interactions between a pair of nodes \cite{Barabasi2004}, \cite{Mcgillivray2018}. There are multiple types of biological interaction networks that represent different biological mechanisms and are based on different types of interaction \cite{Vidal2011}. Many of these biological interactions are publicly available through various specific databases such as KEGG \cite{Kanehisa2012}, Reactome \cite{Matthews2009}, among others. They can be leveraged as DK to deliver a mechanistic and relational inference component which can be integrated to a statistical-probabilistic framework (Fig. \ref{fig:DK_integration_diagram}). 

Pathway-level representations, which represent sets of the pathway genes subsumed into the pathway nodes, with the interactions between the individual genes are also collectively involved in biological processes, such as cell proliferation and death. Thus, malfunction of the pathways can lead to disease. Taking into account the topology of gene interactions as prior knowledge may further help to characterise new genes or disease modules. Many network models have been developed to use known gene-gene interactions for prediction, based on the assumption that interacting genes tend to produce similar phenotypes. New biomarkers discovered by the DL model can be tracked inside the model more easily when the model’s design conforms to biological relations.

The biological pathways can be integrated as curated knowledge on the molecular relation, reaction and interaction networks, covering metabolism, cellular processes, organismal systems, and human diseases and they are widely used to analyse omics data. The pathway construction function can be either a data-driven objective (DDO) or a knowledge-driven objective (KDO) \cite{Viswanathan2008}. The first component is used to establish gene or protein associations identified in a particular experiment. Knowledge-driven pathway construction is associated with the development of a detailed knowledge base for specific areas of interest. There are various approaches to mapping the organisation of cellular functions using molecular interaction networks in which the edges represent interactions between genes, proteins or metabolites. Protein-protein interaction (PPI) data are used to construct networks of reactions important for the regulation and implementation of most biological processes in which proteins have been shown to interact with functionally related proteins. Such an organisation results in the emergence of `functional modules', i.e. functionally related sub-networks in which there is a statistically significant aggregation of nodes with an associated cellular function. Co-expression data, genetic interaction data, and combined data types have been also used to generate similar molecular interaction networks. 

\subsubsection{Data augmentation with domain knowledge}

In this subsection we focus on domain knowledge being used to pre-process the input data in order to change its representation by enrichment or augmentation: from measured omics values as matrices into pathways, networks and graphs (Fig. \ref{fig:DK_integration_diagram}A). First, we discuss how the knowledge of pathways derived from databases was integrated into the model in the reviewed studies. 

\paragraph{At an input level, pathways are mapped to scores, graphs or images.}
Oh et al. \cite{ohPathCNNInterpretableConvolutional2021} demonstrated the method called PathCNN to build an interpretable CNN model using multi-omics data including mRNA expression, copy number variation (CNV) and DNA methylation from the cBioPortal database. Information about pathways together with the associated genes in each omics type is extracted from the KEGG database. Input data at a gene level is converted into pathway-level profiles, and then Principal Component Analysis is applied to extract 3 principal components (PCs). Then, vectors containing PCs of all pathways are represented as a pathway image of a sample (set of pixels) combining all multi-omics data. Images are the input to the CNN model. As an explanation method, Grad-CAM \cite{Selvaraju2017} was used to identify pathways impacting cancer survival predictions by identifying the parts of an image that are most discriminative. The authors assumed that relevant pathways were more likely to be detected if they are grouped together on the pathway images. They managed to highlight the pathways (`pixels') that were of importance for the prediction of long-term survival of glioblastoma patients.

Another model which allows for integration of multi-omics data on pathway level was proposed by Lemsara et al. \cite{lemsaraPathMEPathwayBased2020}. In the multi-modal sparse denoising autoencoder model, multi-omics features are mapped to NCI pathways. Each pathway is represented as a score obtained via autoencoder, then bi-clustering is applied. The model clusters patients based on three-omics data types, including gene expression, miRNA expression, DNA methylation and CNVs data. The SHAP method is used `to understand the impact of individual omics modalities and features on the autoencoded score [...] learned for each pathway' \cite{lemsaraPathMEPathwayBased2020}.

Lee et al. \cite{leeCancerSubtypeClassification2020} proposed a DL model for cancer subtype classification, which used 287 pathways retrieved from KEGG database. Pathways were used to build a graph in which a set of nodes represents genes and a set of edges represents molecular interactions between genes in the pathway. Gene expression profiles from RNA-seq were mapped to nodes represented as a vector. To model each pathway, they used a graph convolutional neural network (GCN), which can capture localised patterns in data and consider interactions among genes. In this way, they built multiple GCNs, one for each of the 287 pathways. Then, a multi-attention based ensemble combines all the pathway models into a single one through two attention levels (pathway-level and ensemble-level). This is followed by a multi-layer perceptron (MLP) for a cancer subtype classification task. The attention mechanism allows for highlighting pathways that are important for the classification, and falls into the ARCH category as notion of pathways directly impacts the model's architecture.  In addition, DK is used POST-HOC to explain the differences between gene expression and interactions between different subtypes in terms of pathways. The authors used the network propagation method on a pathway–PPI network, where the PPI was derived from the BIOGRID database. 

\paragraph{PPI networks as a prevalent type of graph based input.}
An example of a DL model for the integration and analysis of multi-omics data is DeepMOCCA \cite{althubaitiDeepMOCCAPancancerPrognostic2021}. 
DeepMOCCA is a survival prediction model, which integrates DK using PPI networks to transform the input data representation into a graph. The PPI networks are obtained from the STRING database. The multi-omics data is mapped into the nodes, which represent combination of genes, transcripts and proteins. The edges reflect physical and other functional interactions between them. Then, the graph is an input to a GCN with a graph attention mechanism. 
Additionally, as POSTHOC DK integration, cancer driver genes listed from the COSMIC database \cite{COSMICdb} are used to interpret the averaged rank derived from the attention mechanism. By looking at genes with repeatedly high scores across samples but not yet reported as cancer genes, the attention mechanism allows for the generation of new hypotheses. Therefore, DeepMOCCA allows for identification of prognostic markers and cancer driver genes. 
The authors of DeepMOCCA \cite{althubaitiDeepMOCCAPancancerPrognostic2021} also investigated the sample representations in the hidden layer of the network (before the Cox regression) with t-SNE visualization \cite{NIPS2002_6150ccc6} and compared their similarity between cancer types. They suggested that this kind of analysis in reduced dimensional spaces could support patient stratification.

Similarly, Chuang et al. \cite{chuangConvolutionalNeuralNetwork2021} used the PPI network to change the input representation. 
However, their model maps the PPI network into 2D space by using spectral clustering and combines it with the gene expression data to generate images of cancer-related networks of different types of cancer for a CNN model. More specifically, the adjacency matrix (from the PPI network) is reduced to 2 eigenvalues and represented as 2D images. Then a CNN model is trained for cancer type classification. Unfortunately, spectral clustering renders tracing the signal back to individual input features very difficult. This computational step makes significantly reduces the model's interpretability.

Another DL model integrating PPI networks was developed by Chereda et al. \cite{cheredaExplainingDecisionsGraph2021}. They use the PPI network from the HPRD database \cite{periDevelopmentHumanProtein2003, prasadHumanProteinReference2009} to structure the gene expression data. Input data is transformed into a graph and used in GCN model, which is trained to classify expression profiles from breast cancer patients into metastatic or non-metastatic. 
They developed a Graph Layer-wise Relevance Propagation to interpret the outputs of the GCN. They used this explainability method to build a patient-specific subnetwork containing the genes that contribute the most to a prediction. 

Ramirez et al. \cite{ramirezClassificationCancerTypes2020} investigated four models for expression-based cancer type classification (into a cancer subtype or normal tissue gene) using a GCN-based model. The input graphs were generated based on: the co-expression (using Spearman correlation), the co-expression+singleton, the PPI, and the PPI+singleton networks from the STRING database \cite{STRINGdb}. 
As an interpretability method, they use an \textit{in silico} perturbation procedure. Gene expression is successively set to 0 or 1 before passing through the model and examining how the prediction accuracy is affected by this manipulation. The more important for the classification the gene is, the greater the change in accuracy will be observed. This effect is captured with what the authors called a \textit{gene-effect} or \textit{contribution score}, defined as `the larger prediction accuracy change of the labeled cancer type', and calculated for each gene for all classification labels (33 tumor types plus normal).

Schulte-Sasse et al. \cite{Schulte-Sasse2019} combined three omics data types, gene-gene interaction network and PPI network from Consensus Path DB (CPDB). DK was intergrated both in PRE and to assign labels in the dataset. First a gene-gene interaction network is created, where some weak correlations are discarded based on DK from PPI. Such graph is an input to a GCN which is trained to predict whether a gene is associated with the disease or not.
To derive a collection of positive and negative labels for genes in the dataset (\textit{y} - true labels), network of cancer genes (NCG), COSMIC, OMIM and KEGG are used. As the output of the model and true labels depend on the integrated DK, POSTHOC category is also assigned to this model.
The authors demonstrate that including the interaction networks with a GCN classifier helps to classify and predict novel genes as well as entire disease modules. Using the Layer-Wise Relevance Propagation (LRP) \cite{Binder2016}, they are able to dissect which features drive the classification whether a gene is a driver gene or not and to identify, for each gene, neighboring interacting genes that most influence its classification. This results in building sub-modules consisting in a directed graph of gene-gene LRP contributions. As an illustration, this revealed that important neighboring genes of the cancer gene SAPCD2 are enriched for other drivers, suggesting that PPI between these genes are important for the classification.

Liu et al. \cite{liuNetworkbasedDeepLearning2021} developed network-embedding based stratification method (NES). The method constructs the patient vectors based on the network-embedding of the PPI network. More specifically, a struc2vec \cite{Ribeiro_2017} network embedding approach is used. Although this provides relatively good performance in classification of patient subtypes from large-scale patients’ somatic mutation profiles, the method lacks interpretability. The author do not attempt to analyse inner working of the model, which may be due to struc2vec embedding of the input graph, which makes the inference very difficult.

Liu and Xie \cite{liuTranSynergyMechanismdrivenInterpretable2021} developed TranSynergy, to predict the synergistic drug combinations of cancer therapy. Information from the PPI network, gene dependency, and drug-target association are integrated into the model. They proposed a Shapley Additive Gene Set Enrichment Analysis (SA-GSEA) with the aim of deconvoluting `genes that contribute to the synergistic drug combination'. Their SA-GSEA method proceeds by ranking the features (i.e. genes) based on these values and then conducting a gene set enrichment analysis. This approach offers perspective for therapeutic approach and decisions in the context of personalized medicine.

\paragraph{Data enrichment and augmentation driven by relations in the input data.}
Apart from DK extracted explicitly from knowledge bases (e.g. specific pathways), the multi-omics data can be enriched or augmented by using relations derived from the input data, for instance by calculating correlations between gene expression. Studies described below utilise such data enrichment via:  co-expression network, co-expression eigengene matrices, sample similarity networks or guidance graphs with GLUE (graph-linked unified embedding). Of note, expert knowledge is required to define or select appropriate method.

Huang et al. \cite{huangSALMONSurvivalAnalysis2019} proposed SALMON (Survival Analysis Learning with Multi-Omics Neural Networks).
The input to the model consists of mRNA- and miRNA-seq co-expression eigengene matrices. They are derived from lmQCM algorithm \cite{zhangNormalizedImQCMAlgorithm2014}, PRE step. Patient features: diagnosis age, ER and PR status, copy number and tumor mutation burdens are integrated at a later stage.  
The model predicts Cox proportional hazard ratio (survival) for the TCGA breast cancer dataset. As interpretation method, the perturbation procedure measures the importance of each input variable for survival prognosis. Features are ranked according to how much the concordance index (a metric for quantifying how survival prognosis models perform) is decreased. In this POSTHOC interpretation, the authors performed Gene Ontology (GO) and cytoband enrichment from ToppGene Suite to inference the biological implication from the feature ranking.
In this way, Huang et al. \cite{huangSALMONSurvivalAnalysis2019} identified that the diagnosis age and PR status along with five mRNA-seq co-expression modules are the most determinant features. Genes belonging to these leading co-expression modules were further functionally assessed with gene set enrichment analysis. 

A similar way of determining the contribution of input features can be used to identify biomarkers, as illustrated by Wang et al. and their MOGONET model \cite{wangMOGONETIntegratesMultiomics2021}. In MOGONET, DNA methylation, mRNA- and miRNA-seq data are transformed into sample similarity networks. Each network enters a separate GCN. The omic-specific label distributions are then concatenated and integrated with a view correlation discovery network (VCDN), which `can exploit the higher-level cross-omics correlations in the label space' \cite{wangMOGONETIntegratesMultiomics2021}. They identified distinct biomarkers for each of the investigated diseases and performed gene set enrichment analysis yielding results consistent with previous studies.

Another graph embedding of the indput was proposed by Cao and Gao \cite{caoMultiomicsIntegrationRegulatory2021} in a modular framework, called GLUE (graph-linked unified embedding). GLUE utilizes prior knowledge via a knowledge-based graph, called `guidance graph'). 
The method combines omics-specific variational autoencoders with a `guidance graph', which models regulatory interactions across omics layers. The method was used to integrate unpaired single-cell triple-omics data. The nodes in the guidance graph correspond to the features of each omics layer, and edges represent signed regulatory interactions.

Xing et al. \cite{xingMultiLevelAttentionGraph2020} proposed a multi-level attention graph neural network (MLA-GNN) for multi-task prediction. As a first step in the model, the omics data (unimodal, e.g. proteomics or transcriptomics) are converted into a weighted correlation matrix (WGCNA; \cite{Langfelder2008}). Built for the full dataset, the WGCNA represents a coexpression network, from which an edge matrix is derived. Next, a patient-specific graph can be constructed, where the node values are given by the gene expression level in a given sample, and edges between nodes are drawn according to the WGCNA analysis. The graph serves as input to the first (out of three) graph attention layer (GAT) of the DL model. Features from these 3 GAT are then vectorised after a linear projection, and finally fused into a single vector, which finally passes through sequential fully connected layers in the prediction module. Finally, a full-gradient graph saliency (FGS) mechanism is implemented to interpret the predictions. 

\paragraph{Mapping Domain Knowledge as a direct input to DL models.}
The degree to which a gene is essential for cancer cell proliferation is defined as gene dependency \cite{Tsherniak2017}. Chiu et al. \cite{chiuPredictingCharacterizingCancer2021} proposed a DeepDEP autoencoder (AE) to predict gene dependency profile based on the representations learned from high-dimensional genomic data, including DNA mutation, gene expression, DNA methylation, and copy number alteration (CNA). The model includes molecular signatures of the chemical and genetic perturbations from MSigDB as unique functional fingerprints of a gene dependency of interest. First, five AEs (one for each type of input data) are trained on unlabled tumor data, then the outputs from five encoders are combined and passed to DNN. As one of the AEs is trained on fingerprints from MSigDB, which is a DK, we considered the integration as PRE.
Based on DeepDEP, the authors performed detailed post-hoc analysis including input data perturbation, exploration of the latent layers, signature scores and multi-variable linear regression. 

\subsubsection{Explicitly defined architecture}
In this section we discuss DL models that use domain knowledge to modify a standard densely connected DL model's architecture in order to improve both biological plausibility and interpretability (Fig. \ref{fig:DK_integration_diagram}B). 

\paragraph{Pathways are used to define connections.}
Elmarakeby et al. \cite{elmarakebyBiologicallyInformedDeep2021} combined \textit{ex ante} and \textit{ex post} interpretability approaches, proposing a novel neural network architecture - pathway-aware multi-layered hierarchical network (P-NET). It was built using a set of 3,007 curated biological pathways from the Reactome database. The model predicts disease state in prostate cancer patients on the basis of somatic mutations and copy number alterations data. Encoding the relationships that exist in the Reactome dataset focuses the network on interpretability at the design stage (ARCH).

P-NET comprises one layer to encode the genes and five for the pathways. The input layer corresponds to the features that can be quantified and passed through the network. Three nodes from this layer (representing mutations, copy number amplification and copy number deletion) are connected to one node in the subsequent layer. The connections of the second layer reflect gene-pathway relationships whereas those of the next layers are arranged according to parent-child relationships borrowed from Reactome.
For a given patient, the trained NN will return its probability to have metastatic cancer.
For each sample, features can be ranked by importance score in a layer-wise manner using DeepLIFT, where sample-level scores are aggregated to obtain the global importance \cite{Shrikumar2017}. To gain additional insights into the information flow inside P-NET, the authors evaluated how a change in input sample label affects the activation of a node.

Deng et al. \cite{dengPathwayGuidedDeepNeural2020} proposed a pathway-guided deep neural network (DNN) framework to predict drug sensitivity in cancer cells, using known biological signaling pathways, the expression profiles of cancer cell lines, drug - protein interactions, and drug sensitivity datasets. The pathway maps were obtained from the KEGG database. DK was integrated into the DNN model via the layer of pathway nodes and their connections to input gene nodes and drug target nodes. 

Zhao et al. \cite{zhaoDeepOmixScalableInterpretable2021} proposed a scalable, and interpretable DL model, called DeepOmix, for multi-omics data integration and survival prediction. DeepOmix incorporated prior biological knowledge defined by users as the functional module input (such as signaling pathways in this analysis). The pathway gene sets were downloaded from the Molecular Signatures Database (MSigDB) (KEGG and Reactome). DeepOmix integrated multi-omics data as an input gene layer, where nodes of the gene layer are connected with a functional module layer based on the DK. Again, the pathways defined whether there is a connection between nodes. 

Feng et al. \cite{fengInvestigatingRelevanceMajor2021} proposed a DL model, called DeepSigSurvNet, based on a set of (46 selected) signaling pathways from the KEGG database for cancer patients' survival prediction and outcome. The model identifies the individual patterns of these signaling pathways to four types of cancer using gene expression and copy number data (multi-omics data and clinical factors integrated into the model). Not-densely connected layers are followed by CNN with inception modules. For interpretability, Smoothgrad \cite{Smilkov2017} is used to assess how perturbation added to the signaling pathways affects the model's predictions. 
This allows scoring the relevance of each pathway for each cancer type. Then the distributions of the relevance scores of each pathway between different cancer types is compared. The authors noted that striking discrepancies arise among the cancer types and also that for a given cancer type only a small subset of the pathways have high relevance scores. This latter observation could be of interest for prioritising drug or drug combinations that target these driver pathways.

Zhang et al. \cite{Zhang2021} used a DL architecture constrained by the 46 pathways, with a pathway layer that follows the gene layer. Similarly to Feng et al. \cite{fengInvestigatingRelevanceMajor2021}, connections between the two layers are sparse, and connect genes only to pathways to which they belong. They trained the model (`consDeepSignaling') for predicting drug responses in cancer cell lines from the data of dose response and multi-omics (gene expression and copy number). The output from the last layer represents the predicted area under the experimental dose-response curve value of the drug effect on a given cancer cell line. By using Smoothgrad, they analyze the distributions of the importance scores of the signaling pathways from all samples and highlight those important for drug response prediction.

Hao et al. \cite{haoPASNetPathwayassociatedSparse2018} proposed a Pathway-Associated Sparse Deep Neural Network (PASNet) to accurately predict patient prognosis and describe complex biological processes related to prognosis by incorporating curated biological pathways from the MSigDB (Reactome). The sparse DL architecture of PASNet modeled a multilayered, hierarchical biological system of genes and pathways enabling for model interpretability. PASNet included a pathway layer where each node indicates an individual biological pathway (linked with input genes) and a hidden layer which represented hierarchical nonlinear relationships of biological processes into account. The associations between the gene layer and the pathway layer were established by well-known pathway databases (e.g., Reactome and KEGG).

Another sparsely connected DL model is a sparse Variational Autoencoder architecture, VEGA (VAE Enhanced by Gene Annotations) proposed by Seninge et al. \cite{seningeVEGAInterpretableGenerative2021}. The decoder connections are informed by user-provided biological networks based on gene annotation databases (e.g., Reactome). VEGA performance was tested using pathways, gene regulatory networks and cell type marker sets as the gene modules that define its latent space. VEGA was shown to be useful in understanding the response of a population of a specific cell type to a variety of perturbations.

To predict cell states from gene expression profiles, Fortelny and Bock \cite{fortelnyKnowledgeprimedNeuralNetworks2020} proposed Knowledge-Primed Neural Networks (KPNNs) aiming at providing a biologically interpretable DL model.
Their approach combines \textit{ex ante} and \textit{ex post} explainability methods.
The fully connected NNs were replaced by networks derived from prior knowledge of biological networks, including the signaling pathways and gene-regulatory networks. To do this, the authors assumed that most of the regulatory relationships important for the biological system of interest had already been discovered in other contexts. In KPNNs, each node corresponds to a protein or a gene, and each network edge corresponds to a regulatory relationship that has been documented and annotated in biological databases. The model was trained based on single-cell RNA-seq data. Of note, contrary to previously described models,  the KPNN architecture allows for skipping layers.
As for the post-hoc analyses, they focused on the node weights applying a perturbation procedure. It quantifies, for each node, how the addition of small noise is reflected in changes in the outputs. In this way, they evaluated the global importance of the node. These informative weights (in absolute value) can therefore be used to identify likely relevant transcription factors and/or signaling proteins.

\paragraph{Gene Ontology used to define architectural constraints.}
Based on terms extracted from Gene Ontology (GO), the system hierarchy can be structured. Each GO term is associated with a number of genes and gene products, hence genes can be organised into a hierarchy of nested gene sets. Multi-scale hierarchical interactions among biological entities such as GO terms and genes can be encoded as a list of relations. Below, we describe two studies that make explicit integration of GO into ARCH.

The response of cancer cells to therapy depends on biological as well as chemical factors \cite{Turner2015}. To predict drug responses, Kuenzi et al. \cite{kuenziPredictingDrugResponse2020} developed a DL model, called DrugCell, a modular neural network with two branches. The model combines conventional DNN that process compound chemical structures with a Visible Neural Network (VNN) processing binary encodings of individual genotypes. The DrugCell system hierarchy was structured from a literature-curated database. The VNN was guided by a hierarchy of molecular human cell subsystems, taken from 2,086 biological processes from the GO database. In DrugCell, RLIPP \cite{Ma2018} analysis leads to the identification of the gene embedding network subsystems that most contribute to the cell response prediction. Interestingly, Kuenzi et al. \cite{kuenziPredictingDrugResponse2020} further exploited their approach and confirmed the validity of the hypotheses derived from it. Using cell line data, they demonstrated that subsystems identified as important (as evaluated with the RLIPP scores) for the response to a given drug can reveal synergy of drug combination. In addition, they further showed, using patient-derived xenograft models (PDX) data from a public database, how DrugCell can be used to suggest drug combination treatments. DrugCell constitutes a promising example of how analysis of the inner workings of a DL model could translate into therapeutic recommendations.

Another model using GO is Factor Graph Neural Network model proposed by Ma and Zhang \cite{maIncorporatingBiologicalKnowledge2019}. 
Each node in the model corresponds to a biological entity such as genes or GO terms (i.e., gene nodes and GO nodes), which forms a bipartite graph. The model is based on the RLIPP analysis (`relative local improvement in predictive power') and is used to predict tumor stages of kidney and lung cancers and also to classify kidney samples in normal vs. tumor tissues. The method calculating attention matrices allows `capturing multi-scale hierarchical interactions [by assigning] weights to connections between different layers'. By investigating the weights in the last hidden layer, the authors retrieved e.g. the gene ontologies that contribute most to sample classification.

\paragraph{Gene Regulatory Networks used as constraints for VAEs.}
In \cite{shuModelingGeneRegulatory2021}, Shu et al. developed Deep SEM, a VAE-based model which contains a Gene Regulatory Network (GRN) layer in the encoder and Inverse GRN in the decoder. Of note, the weights are shared between these layers. GRN consists of target genes and transcription factors and can be reconstructed based on the representation learnt by the model. DeepSEM is an example of nonlinear mapping from the gene expression to GRN activities. Although no database is used as DK, certainly the GRN layers added to a VAE architecture can be considered as a step forward bio-centric interpretability.
 
\subsubsection{POSTHOC explanations}

Although in previous sections we already described models that use DK both in ARCH and in POSTHOC phases, here we provide examples that integrate DK only for POSTHOC purposes, not impacting the model's design.

A Cox-nnet \cite{chingCoxnnetArtificialNeural2018} is an example of an attempt to link  biological features or functions to the (hidden) nodes of a DNN model solely via POSTHOC analysis. DK is not used in ARCH. Cox-nnet uses a Cox regression as the output layer, extending the Cox-PH model \cite{Therneau2000}. The interpretation of the output includes mapping nodes' weight to regression coefficients, t-SNE, the gene set enrichment analysis with KEGG pathways and computation of partial derivatives of the output. Results from Cox-nnet compared favourably with those from Cox-PH from a biological perspective, revealing for example the importance of the BAI1 gene in the p53 pathway or MAPK1 in several cancer-related pathways. Importantly, POSTHOC interpretation is executed not only via expert (author) evaluation, but systematically using DK about known relations extracted from a database. 

A frequently used POSTHOC interpretation method is the exploration of the association between latent representations with input covariates (e.g. phenotypic features of the patients) \cite{guoDeepLearningbasedOvarian2020}. This approach is of particular interest for models such as autoencoders (AEs) and variational autoencoders (VAEs). In these models the input data is compressed into a reduced (latent) representation and then reconstructed back from the encoded representation with the least possible error. Due to appealing dimensionality reduction abilities, AEs and VAEs are frequently used within the oncology domain (e.g. \cite{10.1093/bioinformatics/btz158, simidjievskiVariationalAutoencodersCancer2019, hiraIntegratedMultiomicsAnalysis2021}). They can be used together with PCA, UMAP \cite{Mcinnes2018}, t-SNE \cite{VanderMaaten2008} or other algorithms \cite{Anowar2021} for data visualization, and various clustering methods can be used on top of that. POSTHOC analyses can then be performed on the weight parameters and/or on the compressed data for gaining biological insights on what the model learned. As an example, XOmiVAE was develop to solve supervised and unsupervised tumour classification tasks \cite{withnellXOmiVAEInterpretableDeep2021}. It uses DeepSHAP explanation \cite{Lundberg2017} to explain novel clusters generated by VAEs. Results are compared with DK derived from i.a. Reactome and GO.

Similarly, Kinalis et al. \cite{kinalisDeconvolutionAutoencodersLearn2019} propose an AE for clustering analysis of scRNA-seq data. They used guided backpropagation (only positive gradients used for the backpass) for computing saliency maps. In their model, saliency values are obtained for each cell and each gene. Gene and gene set importance scores are then computed by averaging across the cells or the corresponding genes, respectively. They use DK in POSTHOC to investigate the latent space of the AE, comparing obtained representation with the pathways (i.e. hematopoietic signatures derived from the DMAP study \cite{novershternDenselyInterconnectedTranscriptional2011}).

In contrast, some AE based models are being developed but no DK is used in PRE, ARCH nor POSTHOC \cite{Way174474, Titus433763, wangExtractingBiologicallyLatent2019, palazzoPancancerSomaticMutation2019}. The architecture proposed by Hira et al. \cite{hiraIntegratedMultiomicsAnalysis2021} can integrate multi-omics data (genomics, epigenomics, transcriptomics). Patient subtyping is obtained first by applying a clustering algorithm on the learned latent features. Clinically relevant latent dimensions are identified by building a univariate Cox proportional hazards (Cox-PH) model for each of them and clustered into survival subgroups. Based on these labels, a Support Vector Machine was trained for allowing survival subgroup classification for new samples. With the aim of identifying biomarkers, a linear model (correlations) is used to map the embeddings of clinical relevance into the gene space.

\section{Discussion}

\subsection{Prevalence of graph representations} 
\begin{figure}[ht!]
\centering
\begin{subfigure}{0.35\textwidth}
  \centering
  \includegraphics[width= \textwidth]{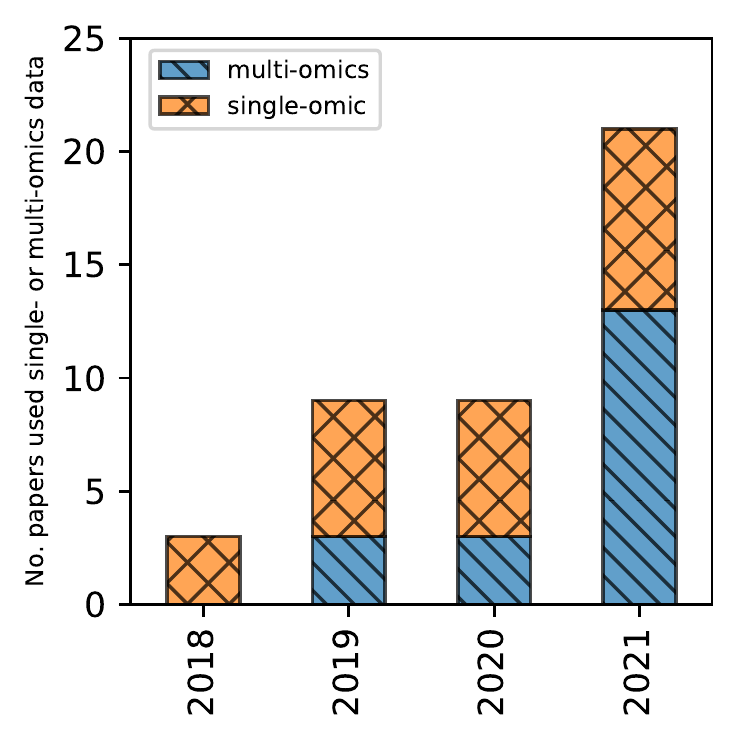}
\caption{}
\label{fig:Task_in_years}
\end{subfigure}%
\begin{subfigure}{0.35\textwidth}
  \centering
  \includegraphics[width= \textwidth]{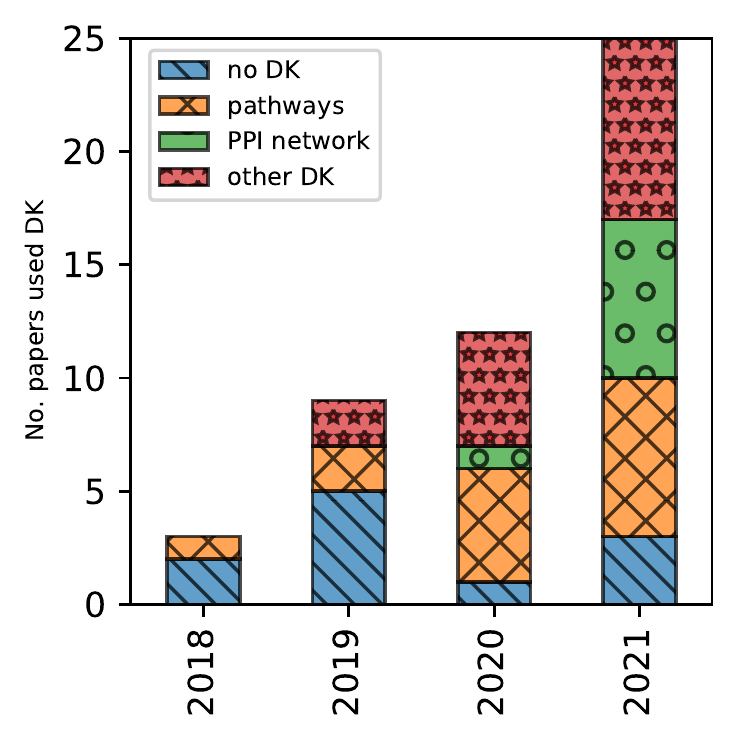}
\caption{}
\label{fig:DK_in_years}
\end{subfigure}%
\\
\begin{subfigure}{.35\textwidth}
  \centering
  \includegraphics[width= \textwidth]{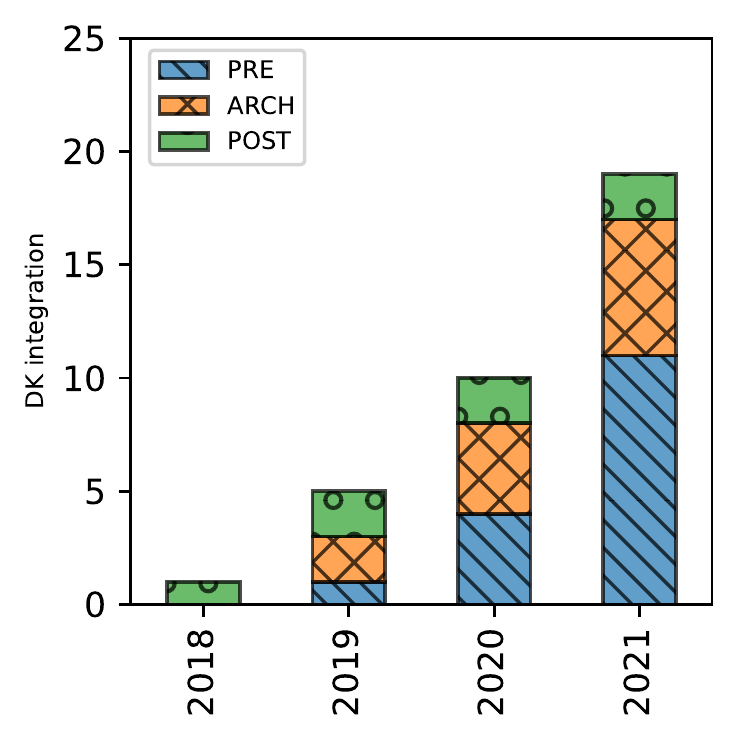}
\caption{}
\label{fig:Repr_in_years}
\end{subfigure}%
\begin{subfigure}{0.35\textwidth}
  \centering
  \includegraphics[width= \textwidth]{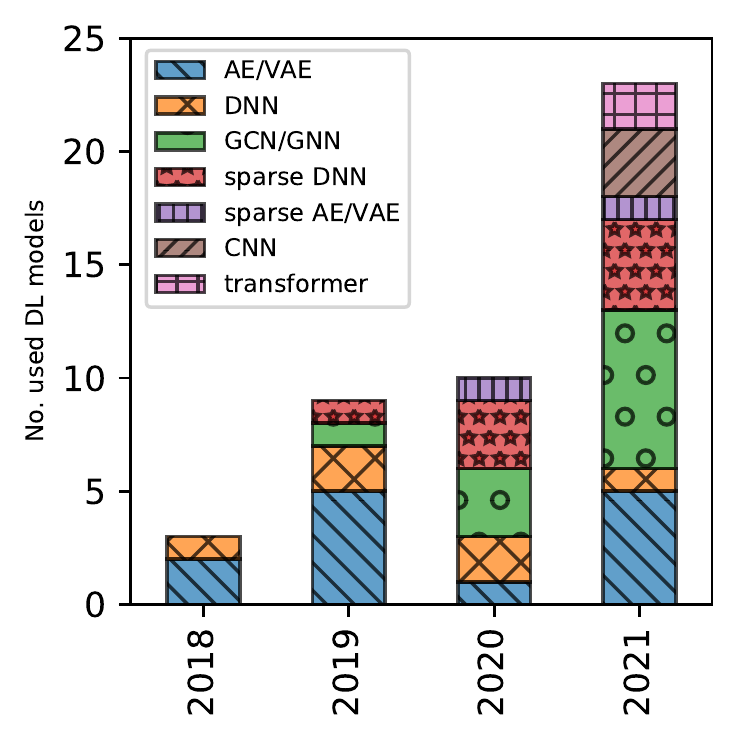}
\caption{}
\label{fig:Models_in_years}
\end{subfigure}%
\caption{The trends in DL models for cancer
There is an upward trend in using multi-omics data (blue) compared to single-omic data (orange) (\textbf{A}) and in the integration of domain knowledge (DK) (orange, green, red) (\textbf{B}) based on recent studies for DL in cancer biology. The most frequently integrated domain knowledge are pathways (orange) and other DK (red) like functional modules with recent increase in the usage of PPI networks. \textbf{C} There are three main categories of DK integration as: input data pre-processing (PRE) (blue), architecture definition (ARCH) (orange) and post-hoc comparison (POST-HOC) (green). There is a trend in the use of DK in PRE step, i.e. DK is used to enrich or augment the input data, which results in a change of data representation; \textbf{D} In recent years, there is an increasing number of specialised DL architectures which encode the structure of biological relations. Graph Neural Networks (GNNs, and Graph Convolution Networks - GCNs) based architectures were the most prevalent used (green). There is an increase in the number of sparse DNN (red) and sparse AE/VAE (blue) models.
}
\label{fig:trends_dk}
\end{figure}

\begin{figure}[ht!]
\centering
\includegraphics[width= 0.8\textwidth]{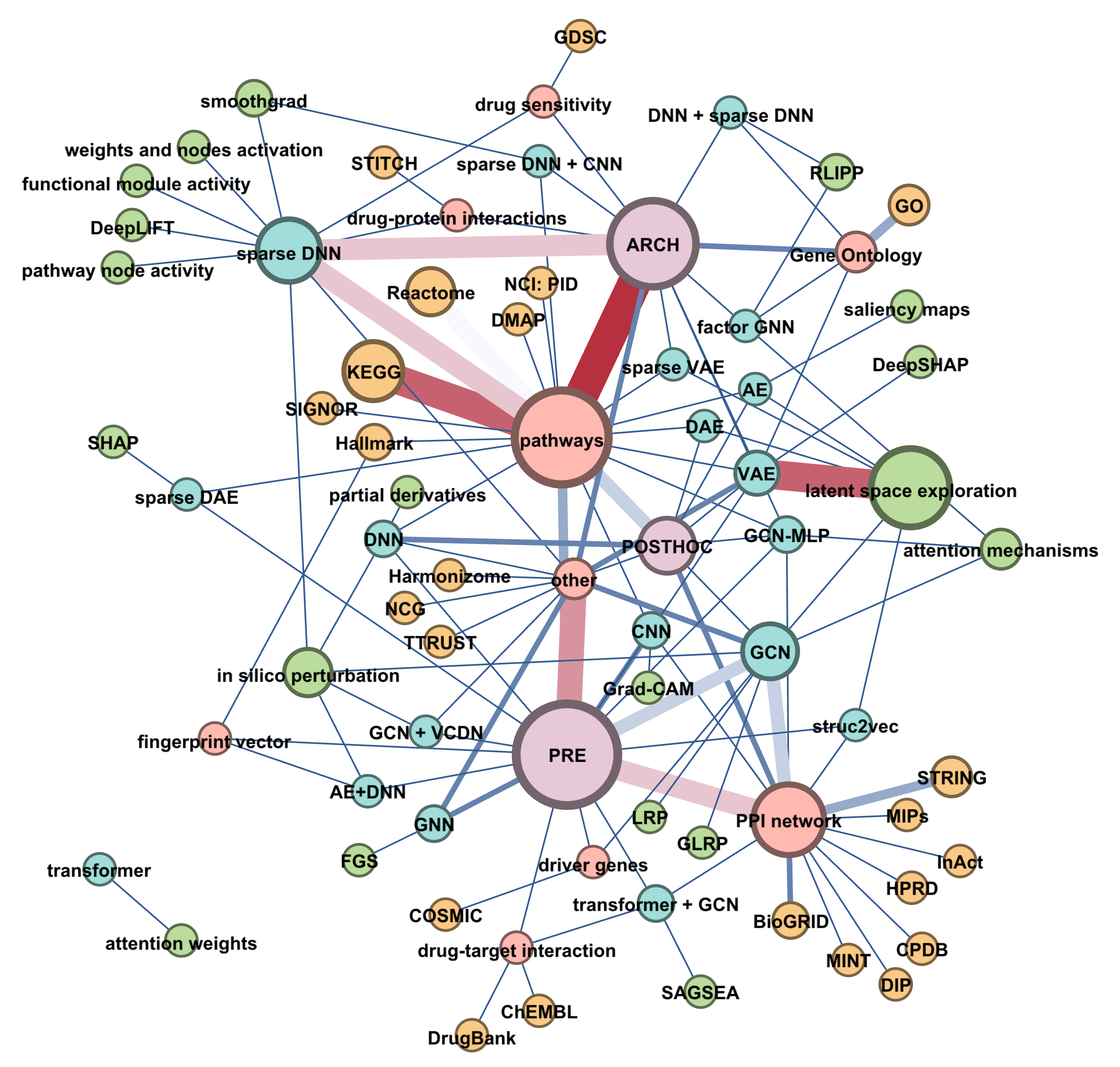}
\caption{Network of relations between key components of the bio-centric interpretability
Network representing the relations between domain knowledge (red nodes), DK databases (orange), DK integration type according to the proposed taxonomy of bio-centric interpretability(purple nodes), DL models (blue nodes) and explainability methods (green nodes). Node size is proportional to the no. occurrences of the entity, edge width is proportional to no. pairs observed in the reviewed papers. We observe strong connections between: ARCH-pathways-sparse DNN; VAE-latent space exploration; PRE-PPI network-GCN; KEGG-pathways.
}
\label{fig:trends_2}
\end{figure}

Recent years have brought an increasing number of specialised DL architectures which encode the structure of biological relations (Fig. \ref{fig:trends_dk}A,B). DL supports non-linear modelling, while encoding complex structures and relationships, in order to learn informative representations at multiple levels of abstraction \cite{Lin2020}. Graph Neural Networks (GNNs, and Graph Convolution Networks - GCNs) based architectures provide a universal support for encoding structural biological knowledge into neural representations. In general, GNNs are a spectrum of models which capture graph dependency by passing interaction between nodes that simultaneously take into account the scale, heterogeneity, and deep topological information of the input data (Fig. \ref{fig:DK_integration_diagram}A). In a biomedical setting, GNNs demonstrate their applicability encoding of topological relations, and mapping them into a high-dimensional embedding space \cite{gao2020}. Compared to other DL models, the advantage of GNN is the ability to integrate relational data into the inference. With the increasing interest in GNNs, we observed a spectrum of new models which combine with explainability methods (Figs. \ref{fig:trends_dk}C,D and \ref{fig:trends_2}).

\subsubsection{Upward trend of graph representations}
Many models were developed to use known gene-gene interactions for prediction, based on the assumption that interacting genes tend to produce similar phenotypes. This resonates with the development in the field of graph neural networks. We observe an increase of GCN/GNN application (1 in 2019, 3 in 2020 and 7 in 2021, (Fig. \ref{fig:trends_dk}D)), which is associated with integration of PPI networks as DK (1 model in 2020, 4 models in 2021, (Fig. \ref{fig:trends_dk}B)).

\subsubsection{32\% of the models which used prior knowledge are GCN models}
GNNs and GCNs models are able to combine heterogeneous omics data types with graph data representations into a predictive model and learn abstract features from both data types. Based on our study, it can be observed that GCNs are the prevalent architectural choice  (Fig. \ref{fig:trends_dk}D). This is due to the fact that the DK is usually represented as a graph (as the phenotype correlates with modules constituting a graph, i.e., sets of related nodes). 

\subsubsection{60\% of the GNN/GCNs used PPIs as a DK}
Due to non-reticular structure data such as graphs, GCNs are successfully used to encode protein-protein interaction networks (PPIs) to predict cancer subtypes, to identify and classify normal tissue and tumour samples for many types of cancer (60\% of the GCNs used PPIs as a DK, Fig. \ref{fig:trends_2}). GCNs can systematically determine which part of the pathway is useful for characterising the tumour. Whether neural networks encoding of biological relations as prior knowledge can accelerate biological discoveries remains largely unknown.

\subsubsection{PPI networks used in PRE to obtain input to GNN}
We observe a pattern that GNN/GCN models are associated with the PPI network application in the pre-processing stage (PRE). Tabular data containing measured multi-omics features are transformed into graphs and then fed into the model. PPI networks are derived from databases such as: STRING, CPDB, HPRD, BioGRID (Fig. \ref{fig:trends_2}).

\subsection{Sparse connections as a key design feature}
\subsubsection{Pathways encoded via sparse connections is an emerging architectural pattern}
We observed an pattern in the approaches towards which employ sparse connections mapping to layers and nodes which have a grounded biological meaning. Domain knowledge integration allows for explicit definition of connections between nodes of DNN. To achieve this, the relational biases of pathways is exploited, where relations are obtained from knowledge bases (KEGG, Reactome, SIGNOR) and used as a mask within the model for removing connections which are not represented. Thus, DK integration in ARCH allows for better, more efficient and meaningful POSTHOC interpretation (Fig. \ref{fig:trends_2}) as well as biological plausibility. As a result, a new architectural paradigm emerges (Fig. \ref{fig:DK_integration_diagram}B), which conforms the architecture to reflect biological relations. 

As the organisation of genes in pathways shapes the high heterogeneity of cancers, taking into account the topology of gene interactions may further help to characterize new gene or disease modules. This is reflected in the prevalence of pathways in DL models: 48.4\% models that integrated DK used pathways (Fig. \ref{fig:trends_dk}B,C). This corresponds to increase in popularity of sparse DNN and sparse AE/VAE models, as the sparsity comes from limited connections between layers defined by the pathways (4 in 2020, 5 in 2021, Fig. \ref{fig:trends_dk}D).

\subsection{Improved support for biomarker discovery}%
From a machine learning (ML) perspective, predicting clinical outcome can be framed as a classification or regression task, and patient or tumor specific subnets can be identified as distinguishing features. However, the high dimensionality of multi-omics data drives an instability in the feature selection process. In this context, stability means that with minor data perturbations, the process is able to preserve the same features \cite{Cun2012, Oller-moreno2021}. Thus, for minor changes in samples, the biomarker detection method should select a consistent/similar gene set. Ideally, the biomarkers can be applied to any sample in the dataset. In general, finding the relevant features remains a major challenge in the high-dimensional, low sample-size setting, in which features are correlated, either by nature (and this is the case in most molecular datasets) or merely by chance (as the number of samples is relatively small). Finding these truly relevant features is significantly more challenging than finding features that provide optimal predictivity. In practice, current algorithms tend to focus on the prediction error of the models and usually are highly unstable, which limits its applicability in a clinical setting and creates barriers for the interpretation of biomedical insight. Stability of the biomarker discovery can be improved by including prior knowledge (i.e. DK) of molecular networks (e.g., pathways or PPI networks; \cite{Oller-moreno2021, Cun2012}). 

\subsection{Research questions - summary}

\subsubsection{What are the perspectives of interpretability across different DL-based frameworks within the cancer research domain?}
 
Based on our proposed taxonomy we argue that to provide biological interpretability to a DL model used in cancer biology, is to enable the domain expert to contemplate the data flow in the entire model and decompose its architectural elements into elements which maps to a biological reasoning and to the structure of the underlying biological mechanisms. We argue that the key explainability property for this class of models is decomposability. Each component can be also viewed as a computational step which transforms the data representation, e.g. in both an explicit or latent form. Although individual computational steps may be mathematically complex, which is inherent to modelling a biological system, they should be organised in the models' architecture in a way that supports the decomposability of the inference process. This will build the representational foundations to deliver \textit{bio-centric interpretability}.

\subsubsection{What are the methods that deliver biological interpretability?}

We argue that a promising category of methods are grounded on sparse connections between neurons (e.g. KPNN), that include skip-connections between hidden layers and that this mechanism supports both bio-centric interpretability and improves the biological plausibility of the inference. Such architecture combined with state-of-the art DL explanation methods allow for tracking back in the network the contribution of biologically grounded components to individual outputs. We argue that designing for bio-centric interpretability, i.e. performing architectural choices which minimise the construction of latent representations which are not easily linked to biological primitives should be at the center of any application of DL for cancer.

\subsubsection{What are the desirable approaches to integration of domain knowledge in the models' architecture?}

DL models can induce a lack of parsimony in data representation (excessive latent features) delivering models which are intrinsically opaque. The application of explainability methods cannot fully circumvent this limitation, limiting the ability of these models to deliver meaningful biological insights. Post-hoc interpretation often leads to confirmation of known existing relations, which is presented as the evidence of the biological plausibility of a model. However, it has been documented that even untrained neural networks can produce saliency maps that appear meaningful \cite{DBLP:journals/corr/abs-1810-03292}. Thus, we argue that bio-centric interpretability may manifest as the ability of the model's architecture to reflect an isomorphism with regard to known biological structures and processes, so that these can be explicitly investigated. Integration of DK allows for the definitions of these architectures. These elements allows for a better use of explainability methods which can rank network components (e.g. nodes activation or edge weights), and return references to biologically grounded elements.

\subsubsection{What are the emerging representation paradigms within these models?}

Based on our Review we identify two main trends:
\begin{itemize}
    \item input data is transformed into graphs or network and fed into GNN or GCN
    \item input data is fed into sparsely connected Deep Neural Network, where connections are defined by biological relations
\end{itemize}

We observed that frequently the multi-omics data is transformed prior to the model input. The transformation extends beyond computational techniques such as the enrichment analysis, and impacts the data representation: tabular data becomes a graph or network (Fig. \ref{fig:DK_integration_diagram}). They can be constructed in data-driven manner, e.g. based on the correlation within the data, like gene-gene interaction networks, or constructed through database DK integration, e.g. input data expressed in nodes and edges of known PPI networks (Fig. \ref{fig:DK_integration_diagram}A). Then, the graph representation is processed in a GNN or GCN. We observe an upward trend in the usage of such models, in most cases using PPI networks as DK.

The second trend focuses more on the architecture of the model, i.e. on the connection between neurons on the network. The input data still can have tabular representation and, because the bio-interpretability comes from carefully crafted architecture, the ability to track back the signal between output and input is not lost. Intuitively, the more times the representation of the data changes in the model, the less interpretable the data flow appears to be. Despite the advantages of graphs in describing biological relations, they might be not the best solution for a DL model, because transforming input data into a graph makes the data flow less transparent (e.g. graph to PCA, then to 2D image; convolutions on graphs). Preserving tabular input data representation may allow for more transparent posthoc explanations, provided that the model's architecture reflects biological relations. For such models, pathways and functional modules derived from knowledge bases are used for defining the the sparse connections (Fig. \ref{fig:DK_integration_diagram}B).

\section{Conclusions}

In this systematic review we focused on the biological interpretability of Deep Learning models that target omics data developed in the domain of cancer biology.  We introduced the new concept of \textit{bio-centric interpretability} and defined its key properties and components. According to a taxonomy centered around this notion, we critically reviewed recent studies in the context of model architecture, domain knowledge integration and biological interpretability methods. 

We found that the convergence between the use of external domain knowledge and the design of architectures which reflect the structure of known biological mechanisms can deliver: (i) the model explainability required by domain experts, (ii) the improvement of the biological plausibility of these models, (iii) the improvement of the explanation quality delivered by post-hoc methods and more fundamentally (iv) the repositioning of DL models from opaque pure-predictors to explainable models which can support new biological insight. The two most common approaches to incorporate DK into the model are to use pathways or PPI networks (Figs. \ref{fig:DK_integration_diagram} and \ref{fig:trends_dk}). They can be used for (i) data augmentation, (tabular mRNA data → graphs based on gene interactions) and (ii) to biologically ground the architecture of the model (e.g. mapping the connections between nodes). Domain knowledge is most frequently represented as pathways and PPI networks, which are derived from public databases, such as KEGG and Reactome, exploiting the existing curated biological knowledge. The vast majority of reviewed models attempt to interpret the output by post-hoc analyses, with a clear pattern: the more domain knowledge is reflected in the model design, the more interpretable is the post-hoc analysis. 
Although expert knowledge is always required to interpret the results, we assert that only the integration of explicit domain knowledge in the model design may lead to the improvement in understanding the underlying biological mechanisms. As the notion of \textit{biological interpretability} is still largely unformalised, we highlight the need for universal bio-centric interpretable methods, so the developed methods are less problem- or application-specific. 

\section{Methods}
In this systematic review, we summarise emerging DL models in cancer biology covering the representation of biological processes, diagnosis and prognosis, and recent progress in biologically informed models. To this end, we started by searching electronic bibliographic databases (PubMed and Web of Science) for relevant studies published between Jan 1, 2018, and Jan 1, 2022. We used the following terms: \textit{multi-omics} and \textit{deep learning} or \textit{computer science} or \textit{neural networks} or \textit{network analysis} or \textit{machine learning} and \textit{cancer} or \textit{cancer biology}. The same search was repeated just before the final experimental analysis for completeness (Mar 1, 2022). 

We concentrated on deep learning methods applied to cancer or at least to those that are linked to straightforward applications in cancer biology using multi-omics data conducted in humans or human cell lines. Furthermore, we also searched the reference lists of published trials and the relevant review articles. At last we only concentrated on DL applications for omics data including: genomics, transcriptomics and epigenomics data from cancer in humans. We excluded studies published in languages other than English, studies with insufficient data (i.e., studies where full texts were not available or irrelevant studies), case reports, editorial materials, comments and meeting abstracts. Similarly, all pre--clinical studies conducted either in animal cell lines or murine models, review articles, meta-analyses or studies performed on animals and animal cell lines were excluded. Papers providing methods that are not directly linked to cancer and functional analysis/insights on biological processes were excluded. Papers centered around medical imaging were excluded (e.g. histopathology and computed tomography) as well papers provided models based on DL and ML using clinical/laboratory data alone. Studies based on microarray data or developed a sequence-based algorithmic framework were excluded as well.

Using the search strategy, we obtained titles and abstracts of retrieved studies and imported them to an endnote. Two authors independently screened identified studies on the basis of prespecified inclusion criteria. All potentially relevant articles were read in full text and a list of eligible studies was created. Data were manually extracted using a structured template and any disagreements were resolved by mutual agreement between these two authors during the process of screening and data extraction, or by intervention of a third author. A standardised data extraction form was used to extract the following fields: authors’ names, year of publication, type of omic data, model's output, type of prior knowledge, prior knowledge databases, type of domain knowledge integration, type of deep learning model/architecture and interpretability method used. 

The multi-omics data can be represented in various ways in the subsequent components of the model. This representation can be changed into a another representation in a series of computations steps. It is crucial to understand these representations, how they are transformed, and how to communicate such transformation during post-hoc inference. We distinguished four bio-centric interpretability components: the integration of different data modalities, the schema level representation of the model, the integration of domain knowledge, post-hoc explainability methods. We took the concept of interpretability and distinguished three categories: architecture-centric interpretability, output-centric interpretability, and post-hoc evaluation of biological plausibility. The association of the model to each of the identified groups was done manually based on the authors' expertise. 

The selection criteria resulted in 42 studies\footnote{Each of the selected studies is from the existing state of the art and not performed by any of the authors.}. We elaborate on the components involved in bio-centric interpretability within DL models, focusing on emerging architectural and methodological advances, the encoding of biological domain knowledge and the integration of explainability methods. The dimensions of bio-centric interpretability for recent studies are presented in Table 1.

\section{Acknowledgements}
Not applicable

\section{Funding}
This project has received funding from the European Union's Horizon 2020 research and innovation programme under grant agreement No 965397.
This project has also be supported by funding from the digital Experimental Cancer Medicine Team, Cancer Biomarker Centre, Cancer Research UK Manchester Institute P126273. 

\begin{figure}[htb!]
\centering
\includegraphics[width= .8\textwidth]{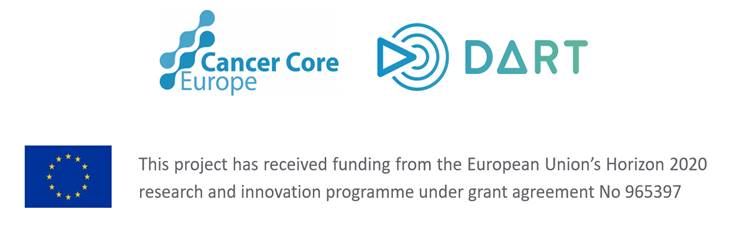}
\captionsetup{labelformat=empty}
\caption{}
\label{fig:cce_dart}
\end{figure}

\section{Competing interests}
The authors declare that they have no competing interests.

\pagebreak

\printbibliography
\end{document}